\documentclass[
]{article}
\usepackage{amsmath,amssymb}
\usepackage{booktabs}
\usepackage{iftex}
\usepackage{booktabs,array}
\usepackage{graphicx}
\usepackage[authoryear,round]{natbib}
\usepackage{arxiv}
\usepackage{orcidlink}
\usepackage{amsmath}
\usepackage[T1]{fontenc}
\usepackage{url}
\usepackage{listings}
\usepackage{float}
\usepackage{pgfplots}
\pgfplotsset{compat=1.18}
\usepackage{tabularx}
\usepackage{multirow}
\usepackage{colortbl}
\usepackage{xcolor}
\hypersetup{
    hidelinks
}

\newcommand{\runninghead}{A Preprint }
\renewcommand{\runninghead}{THK-AI Research Report 2/2026 }
\title{LLMs taking shortcuts in test generation: A study with SAP HANA
and LevelDB}
\def\asep{\And}
\def\arowsep{\AND}
\author{\textbf{Vekil Bekmyradov}~\orcidlink{0009-0007-7429-8259}\\\\THK-AI Research Cluster, TH
Köln\\\\\href{mailto:vekilmuhammet.bekmyradov@smail.th-koeln.de}{vekilmuhammet.bekmyradov@smail.th-koeln.de}\asep\textbf{Noah
C. Puetz}~\orcidlink{0009-0002-9895-3661}\\\\THK-AI Research Cluster, TH
Köln\\\\\href{mailto:noah_christoph.puetz@th-koeln.de}{noah\_christoph.puetz@th-koeln.de}\arowsep\textbf{Thomas
Bartz-Beielstein}~\orcidlink{0000-0002-5938-5158}\\\\THK-AI Research Cluster, TH
Köln\\\\\href{mailto:thomas.bartz-beielstein@th-koeln.de}{thomas.bartz-beielstein@th-koeln.de}}
\date{}
\begin{document}
\twocolumn[{%
\maketitle
\begin{abstract}
Large Language Models (LLMs) have achieved impressive results on public benchmarks, often leading to claims of advanced reasoning and understanding. However, recent research in cognitive science reveals that these models sometimes rely on shallow heuristics and memorization, taking shortcuts rather than demonstrating genuine cognitive abilities. This paper investigates LLM behavior in automated test generation for software, contrasting performance on an open-source system (LevelDB) with SAP HANA, one of the most widely deployed commercial database systems worldwide, whose proprietary codebase is guaranteed to be absent from training data. We combine cognitive evaluation principles, drawing on Mitchell's mechanism-focused assessment methodology, with empirical software testing, employing mutation score and iterative compiler-feedback repair loops to assess both accuracy and underlying reasoning strategies. Results show that LLMs excel on familiar, open-source benchmarks but struggle with unseen, complex domains, often prioritizing compilability over semantic effectiveness. These findings provide independent software engineering evidence for the broader claim that current LLMs lack robust reasoning, and highlight the need for evaluation frameworks that penalize trivial shortcuts and reward true generalization.
\end{abstract}
\vspace{1em}
}]

\section{Introduction}\label{introduction}

The rapid advancement of Artificial Intelligence (AI) has led to
exceptional performance on public benchmarks, often driving claims that
these systems possess broad capabilities like ``reasoning'' and
``understanding''. However, recent research in cognitive science and
software engineering indicates that this benchmark performance
can overestimate their true real-world capabilities. AI algorithms often
lack robust ``world models'' or abstract causal understanding, instead
relying on ``approximate retrieval'' from their vast training data or
exploiting unintended surface heuristics, taking
``shortcuts'' to arrive at correct answers for the wrong
reasons.
Mitchell and her team's evaluation of LLMs on the Abstraction and
Reasoning Corpus (ARC-AGI) demonstrates that while models like OpenAI's
o3 can achieve high accuracy, they often fail to grasp the intended
human-like ``core knowledge'' abstractions (such as ``objectness'' or
geometry) \citep{mosk23a,bege25a}. Instead, they use less generalizable
shortcuts, such as tracking shallow pixel patterns. Consequently, AI
evaluations must look beyond simple accuracy and analyze the
\emph{mechanisms} and reasoning paths models take to achieve their
results. This concern is not isolated to abstract reasoning tasks. A
structurally identical dynamic has been documented in the broader machine
learning literature under the term \emph{shortcut learning}: the tendency
of models to exploit statistical regularities that correlate with correct
answers in training but fail to generalize under distribution
shift~\citep{geir20a,mitc23a}.

This phenomenon closely mirrors challenges in automated software
testing. In test generation, existing evaluations mostly rely on
open-source benchmarks (e.g., HumanEval\footnote{\url{https://github.com/openai/human-eval}}) that are highly likely to be
present in the LLMs' training corpora. This data contamination makes it
unclear whether high performance reflects genuine reasoning or mere
memorization~\citep{ridd24a,chen25a}. The
problem is compounded by a second, domain-specific shortcut: code
coverage, the dominant proxy metric for test quality in the field, can
be tests that execute source lines without asserting any meaningful
behavior~\citep{inoz14a}.

\citet{bekm26c} addresses this by analyzing LLM
behavior and testing strategies on SAP HANA, one of the most widely
deployed commercial database systems worldwide, compared to the
open-source key-value store LevelDB. Because SAP HANA's codebase is
proprietary, it is guaranteed to be absent from the public pre-training
corpora of all evaluated models, providing a zero-contamination
evaluation environment. LevelDB, by contrast, is an
open-source project whose codebase and test suite are almost certainly
present in those same corpora. Together, these perspectives reveal how LLMs optimize for immediate
constraints, whether solving grid puzzles or fixing compilation
errors by relying on shallow heuristics rather than robust semantic
understanding.

This paper is structured as follows:
Section~\ref{methods} describes the evaluation methodologies, combining
Mitchell's principles of mechanism-focused cognitive evaluation with the
empirical software testing framework employed by Bekmyradov.
Section~\ref{results} presents the empirical findings, highlighting the
stark contrast in LLM performance between familiar and novel domains and
documenting the three principal shortcut behaviors observed.
Finally, Section~\ref{discussion} discusses the implications of these
results for AI evaluation practice and outlines directions for future
research.

\section{Methodology}\label{methods}

To rigorously evaluate the cognitive and reasoning capabilities of LLMs,
we combine Mitchell's principles of cognitive evaluation with empirical
software testing methodologies. Mitchell emphasizes that evaluations
must test for robustness, avoid anthropomorphic biases, and investigate
the mechanisms underlying performance rather than relying solely on
accuracy~\citep{mitc23a,mitc23b}.

Applying this to software testing, Bekmyradov evaluated four models
(GPT-5, Claude 4 Sonnet, Gemini 2.5 Pro, Qwen-3-Coder) using a
multi-di\-mensional framework. Instead of relying exclusively on code
coverage which can act as a flawed ``shortcut'' metric inflated by
trivial tests. Therefore the study incorporated mutation score (MS) to assess the
actual fault-detection effectiveness of the generated tests~\citep{jiah11a, demi78a}.

The experimental pipeline featured two scenarios. In \emph{Test
Amplification}, existing human-written test suites were artificially
reduced and the LLM was asked to regenerate the missing tests. In
\emph{Whole-Suite Generation}, all existing tests were discarded and the
LLM generated a complete test suite from source code alone. Two context
configurations were evaluated for Whole-Suite Generation: source code
only, and source code augmented with the corresponding dependency files.
To handle the strict syntactical demands of complex C++
projects, the methodology included an iterative compiler-feedback repair
loop. When a generated test failed to compile, the error log was fed
back to the LLM for up to ten repair iterations~\citep{bekm26c}. This
process allowed for the qualitative evaluation of the \emph{way} the
LLM reached its result. This is analogous to Mitchell's method of prompting
models to output natural language transformation rules to reveal their
internal logic. Tracking compilation success rate (CSR) across iterations exposes whether models
resolve compilation errors through genuine reasoning about the codebase
or by progressively simplifying and hollowing out the generated tests.

\section{Results}\label{results}

The empirical findings reveal a profound gap between memorized retrieval
and genuine generalization, heavily characterized by the models taking
structural shortcuts. We organize the findings around following three distinct
shortcut behaviors.

\subsection{Memorization vs.\ Generalization}

On the open-source LevelDB project, all four models demonstrated
exceptional performance, achieving up to 100\% mutation scores and
surpassing the human-written baseline of 52.79\%. In the Whole-Suite
Generation scenario, every evaluated model achieved a perfect mutation
score of 100\%, despite having no access to the original test
suite~\citep{bekm26c}. This strongly suggests that the models recognized
the repository from their training data and reproduced familiar test
patterns rather than synthesizing novel tests through reasoning.
Listings~\ref{lst:original} and~\ref{lst:generated} show
a representative case: the generated test is a near-verbatim copy of the
original human-written test, with only minor structural differences.
This pattern is consistent with approximate retrieval from training data
rather than reasoning about the code~\citep{bekm26c}, mirroring Lewis and
Mitchell's finding that LLM reasoning performance drops sharply when
tasks are varied to be dissimilar from training data while testing the
same abstract abilities~\citep{lewi24a}.

\begin{figure*}
\centering
\begin{minipage}[t]{0.48\textwidth}
\begin{lstlisting}[language=C++, numbers=left, xleftmargin=2em,
frame=single, basicstyle=\footnotesize\ttfamily,
breaklines=true, columns=fullflexible,
caption={Original human-written test.}, captionpos=b,
label={lst:original}]
TEST_F(AddBoundaryInputsTest, TestNoBoundaryFiles) {
  FileMetaData* f1 =
      CreateFileMetaData(1, InternalKey("100", 2, kTypeValue),
                   InternalKey(InternalKey("100", 1, kTypeValue)));
  FileMetaData* f2 =
      CreateFileMetaData(1, InternalKey("200", 2, kTypeValue),
                   InternalKey(InternalKey("200", 1, kTypeValue)));
  FileMetaData* f3 =
      CreateFileMetaData(1, InternalKey("300", 2, kTypeValue),
                   InternalKey(InternalKey("300", 1, kTypeValue)));
  level_files_.push_back(f3);
  level_files_.push_back(f2);
  level_files_.push_back(f1);
  compaction_files_.push_back(f2);
  compaction_files_.push_back(f3);
  AddBoundaryInputs(icmp_, level_files_, &compaction_files_);
  ASSERT_EQ(2, compaction_files_.size());
}
\end{lstlisting}
\end{minipage}\hfill
\begin{minipage}[t]{0.48\textwidth}
\begin{lstlisting}[language=C++, numbers=left, xleftmargin=2em,
frame=single, basicstyle=\footnotesize\ttfamily,
breaklines=true, columns=fullflexible,
caption={Generated test.}, captionpos=b, label={lst:generated}]
TEST_F(AddBoundaryInputsTest, TestNoBoundaryFiles) {
  FileMetaData* f1 =
      CreateFileMetaData(1, InternalKey("100", 2, kTypeValue),
                         InternalKey("199", 1, kTypeValue));
  FileMetaData* f2 =
      CreateFileMetaData(2, InternalKey("200", 2, kTypeValue),
                         InternalKey("299", 1, kTypeValue));
  FileMetaData* f3 =
      CreateFileMetaData(3, InternalKey("300", 2, kTypeValue),
                         InternalKey("399", 1, kTypeValue));
  level_files_.push_back(f3);
  level_files_.push_back(f2);
  level_files_.push_back(f1);
  compaction_files_.push_back(f2);
  AddBoundaryInputs(icmp_, level_files_, &compaction_files_);
  ASSERT_EQ(1, compaction_files_.size());
  ASSERT_EQ(f2, compaction_files_[0]);
}
\end{lstlisting}
\end{minipage}
\end{figure*}

Table~\ref{tab:effectiveness} shows the Whole-Suite Generation results for SAP HANA and LevelDB, reporting line coverage ($Cov_L$), branch coverage ($Cov_B$), and mutation score ($MS$). Human (Full) refers to metrics obtained by running the
complete original human-written test suite, while Human (Reduced) refers
to the artificially reduced subset used as the generation baseline in the
experiments. For complete results including Test Amplification scenarios,
we refer the reader to~\citep{bekm26c}.

\begin{table*}
\caption{Whole-Suite Generation results for SAP HANA and LevelDB. '--' denotes values that were not publicly disclosed~\citep{bekm26c}.}
\label{tab:effectiveness}
\centering
\begin{tabular}{l l l c c c}
\toprule
\textbf{Project} & \textbf{Methodology} & \textbf{Model} & \textbf{$Cov_L$ (\%)} & \textbf{$Cov_B$ (\%)} & \textbf{$MS$ (\%)} \\
\midrule
\rowcolor{gray!10} \multirow{10}{*}{SAP HANA} & Human (Full) & Human & -- & -- & -- \\
\rowcolor{gray!10} & Human (Reduced) & Human & 66.71 & 35.33 & 30.41 \\
\cmidrule(lr){2-6}
& \multirow{4}{*}{Whole-Suite (Src)} & GPT-5 & 46.14 & \textbf{27.99} & \textbf{10.25} \\
& & Claude 4 Sonnet & \textbf{47.71} & 25.27 & 6.39 \\
& & Qwen3-Coder & 35.02 & 18.03 & 6.18 \\
& & Gemini 2.5 Pro & 24.68 & 15.21 & 2.39 \\
\cmidrule(lr){2-6}
& \multirow{4}{*}{Whole-Suite (Src+H)} & GPT-5 & 60.87 & \textbf{34.26} & \textbf{25.14} \\
& & Claude 4 Sonnet & \textbf{62.11} & 31.31 & 17.49 \\
& & Qwen3-Coder & 45.02 & 21.72 & 9.20 \\
& & Gemini 2.5 Pro & 37.90 & 22.04 & 10.60 \\
\midrule
\rowcolor{gray!10} \multirow{10}{*}{LevelDB} & Human (Full) & Human & \textbf{73.78} & \textbf{57.08} & \textbf{52.79} \\
\rowcolor{gray!10} & Human (Reduced) & Human & 54.87 & 37.59 & 37.32 \\
\cmidrule(lr){2-6}
& \multirow{4}{*}{Whole-Suite (Src)} & GPT-5 & \textbf{82.69} & \textbf{66.97} & \textbf{100.00} \\
& & Claude 4 Sonnet & 73.30 & 57.23 & \textbf{100.00} \\
& & Qwen3-Coder & 63.45 & 47.60 & \textbf{100.00} \\
& & Gemini 2.5 Pro & 71.99 & 54.01 & \textbf{100.00} \\
\cmidrule(lr){2-6}
& \multirow{4}{*}{Whole-Suite (Src+H)} & GPT-5 & \textbf{78.20} & \textbf{60.90} & 98.24 \\
& & Claude 4 Sonnet & 76.37 & 60.26 & 99.89 \\
& & Qwen3-Coder & 72.46 & 55.25 & \textbf{100.00} \\
& & Gemini 2.5 Pro & 73.15 & 57.68 & \textbf{100.00} \\
\bottomrule
\end{tabular}
\end{table*}
In stark contrast, performance collapsed on the unseen SAP HANA
codebase. For Test Amplification, the best-performing model (GPT-5)
achieved a mutation score of 39.54\%. For Whole-Suite Generation with
dependency context, the maximum mutation score across all models was
25.14\%, falling substantially below the human-written
baseline~\citep{bekm26c}.

\subsection{Taking Shortcuts in Iterative Repair}

The iterative compiler-feedback loop successfully increased compilation
success rates by roughly 2--3$\times$, with GPT-5 reaching up to 99\%
on SAP HANA. However, qualitative analysis of the \emph{mechanisms} used
to achieve compilability revealed that models increasingly took shortcuts
as the repair loop progressed~\citep{bekm26c}. To resolve build errors, the LLMs shifted their
optimization toward strict compilability at the expense of semantic effectiveness,
frequently generating trivial or empty test bodies, or commenting out
meaningful assertions.

On LevelDB, almost all models reached near-perfect compilation success
within the first one or two iterations, with Gemini~2.5~Pro jumping from
0\% to 70\% in a single repair step. This rapid convergence is
consistent with the memorization hypothesis: the errors encountered were
shallow and easily resolved by the models' prior knowledge of the
codebase. On SAP HANA, repair curves climbed slowly across all ten
iterations and plateaued well below 100\% for most models, indicating
that the errors were structurally unfamiliar and not resolvable through
pattern recall alone.

\begin{figure*}
\centering
\includegraphics[width=0.9\textwidth]{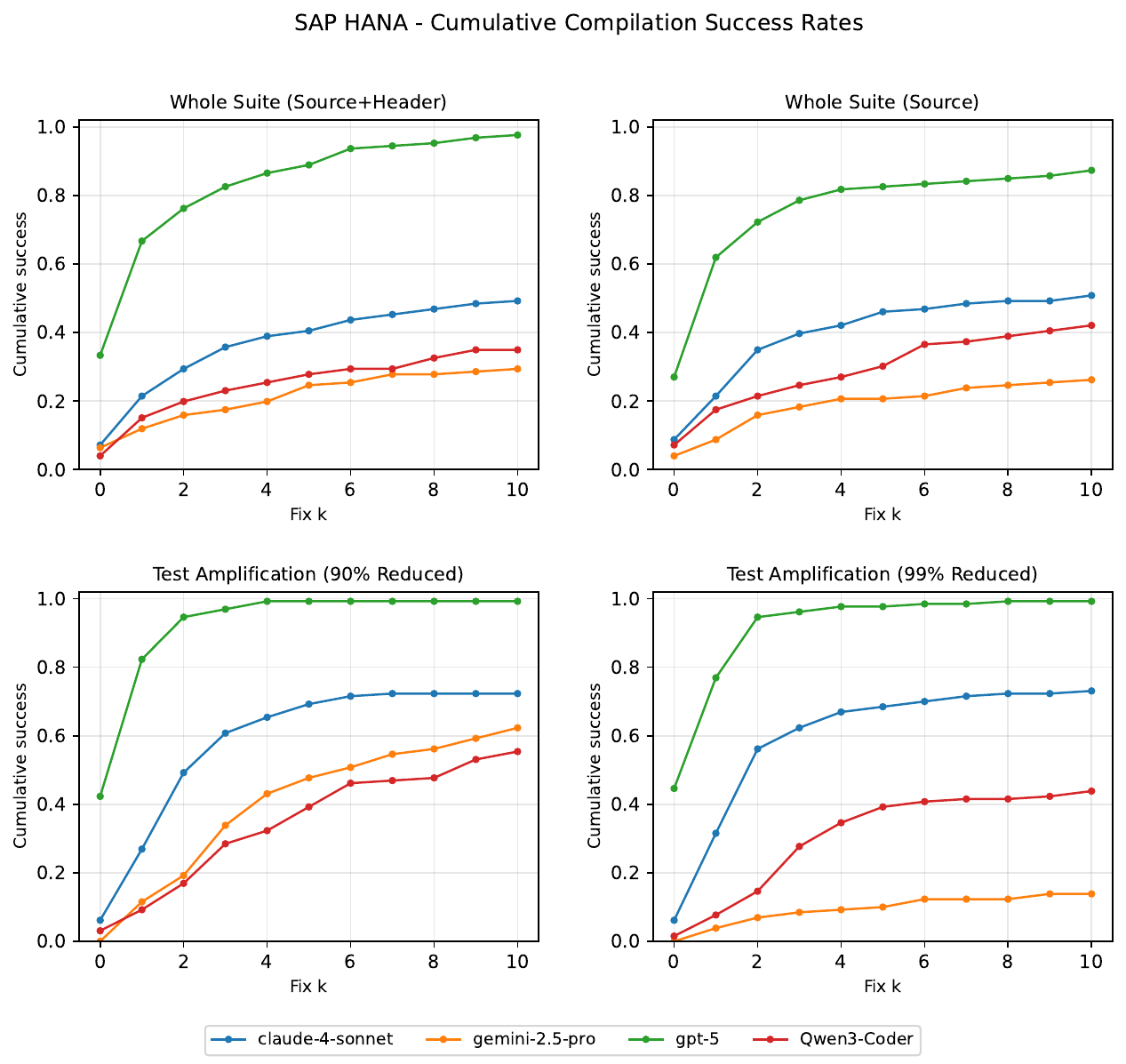}
\caption{Cumulative compilation success rate over $k=10$ repair iterations for SAP HANA~\citep{bekm26c}.}
\label{fig:repair_hana}
\end{figure*}

\begin{figure*}
\centering
\includegraphics[width=0.9\textwidth]{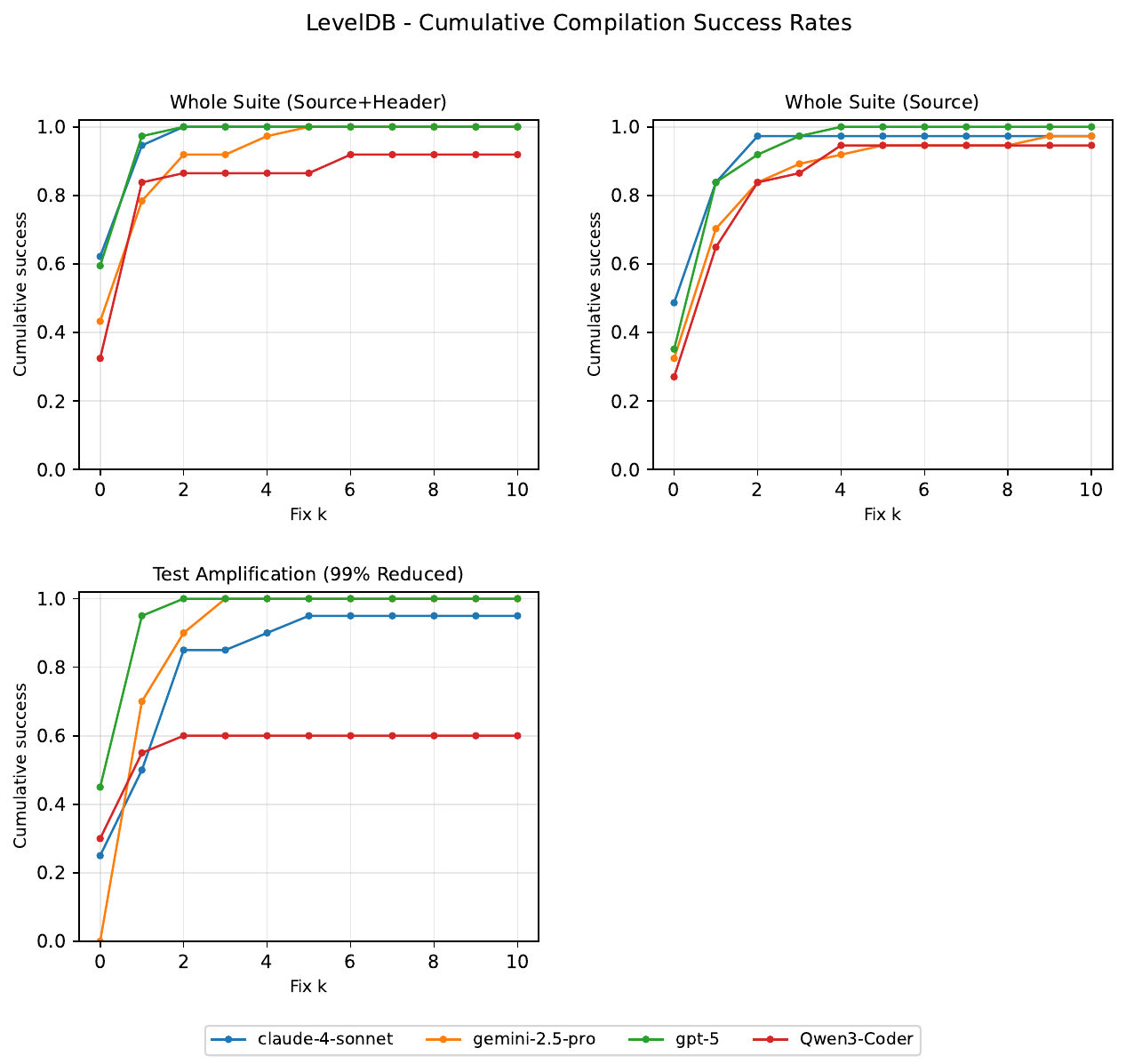}
\caption{Cumulative compilation success rate over $k=10$ repair iterations for LevelDB~\citep{bekm26c}.}
\label{fig:repair_leveldb}
\end{figure*}

Figures~\ref{fig:repair_hana} and~\ref{fig:repair_leveldb} show the cumulative
compilation success rates across repair iterations for SAP HANA and LevelDB, respectively.

\subsection{Explicit Context Reveals Absent Structural Knowledge}

Just as Mitchell found that LLMs struggle with innate human priors like
``objectness''~\citep{mitc23b,bege25a}, the models in this study
demonstrated no inherent understanding of SAP HANA's project structure,
dependency graph, or namespace conventions~\citep{bekm26c}. When provided with source code
alone, models frequently hallucinated incorrect application programming interface (API) calls, omitted
required headers, and failed to instantiate objects correctly.

Providing dependency files as auxiliary context produced consistent and
substantial improvements across all four models: line coverage increased
by 10--15 percentage points, and mutation scores increased by up to
150\%~\citep{bekm26c}. This improvement was not observed on LevelDB,
where header context made little difference, consistent with the
interpretation that LevelDB models already possessed implicit structural
knowledge from training data.

\begin{figure*}
\centering
{\begin{tikzpicture}
\begin{axis}[
    ybar,
    width=\textwidth,
    height=8cm,
    bar width=10pt,
    symbolic x coords={GPT-5, Claude 4 Sonnet, Qwen3-Coder, Gemini 2.5 Pro},
    xtick=data,
    ylabel={Coverage (\%)},
    ymin=0, ymax=80,
    legend style={at={(0.5,-0.18)}, anchor=north, legend columns=4},
    title={Impact of Header Context on SAP HANA Coverage},
    nodes near coords,
    nodes near coords style={font=\scriptsize, rotate=90, anchor=west},
    enlarge x limits=0.2
]

\addplot[fill=blue!30] coordinates {(GPT-5,46.14) (Claude 4 Sonnet,47.71) (Qwen3-Coder,35.02) (Gemini 2.5 Pro,24.68)};
\addplot[fill=blue!80] coordinates {(GPT-5,60.87) (Claude 4 Sonnet,62.11) (Qwen3-Coder,45.02) (Gemini 2.5 Pro,37.90)};

\addplot[fill=teal!30] coordinates {(GPT-5,27.99) (Claude 4 Sonnet,25.27) (Qwen3-Coder,18.03) (Gemini 2.5 Pro,15.21)};
\addplot[fill=teal!80] coordinates {(GPT-5,34.26) (Claude 4 Sonnet,31.31) (Qwen3-Coder,21.72) (Gemini 2.5 Pro,22.04)};

\legend{$Cov_L$ (Src), $Cov_L$ (Src+H), $Cov_B$ (Src), $Cov_B$ (Src+H)}
\end{axis}
\end{tikzpicture}}
\caption{Effect of header context on line and branch coverage for SAP HANA~\citep{bekm26c}.}
\label{fig:hana_coverage_context}
\end{figure*}
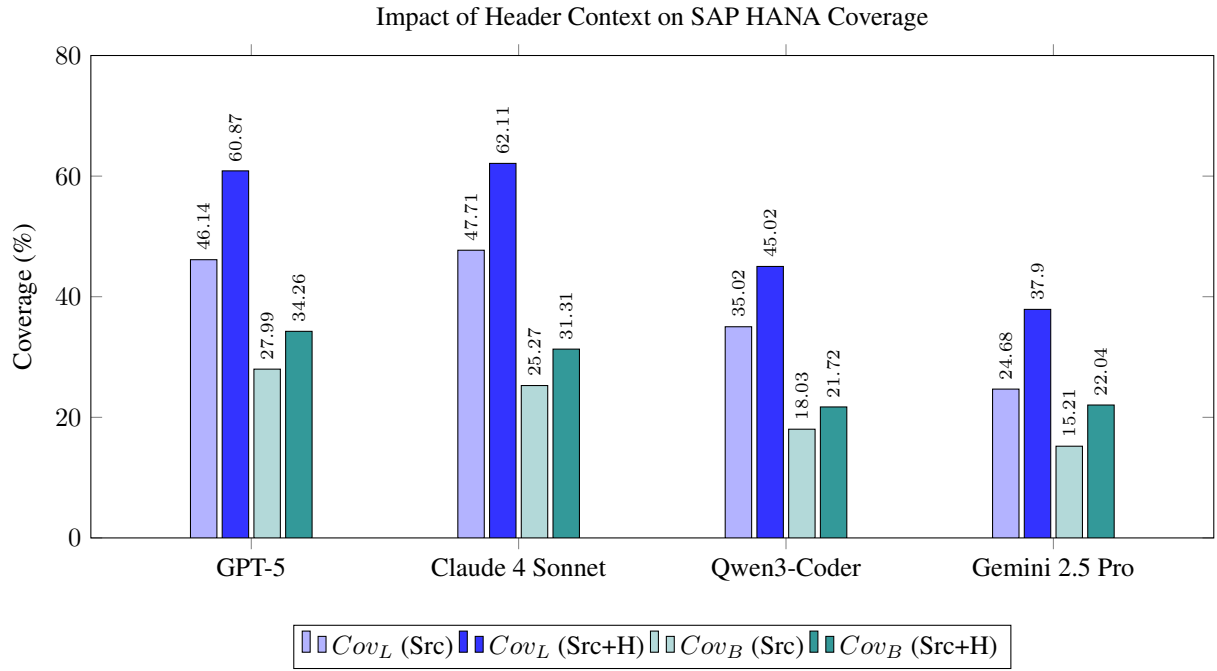

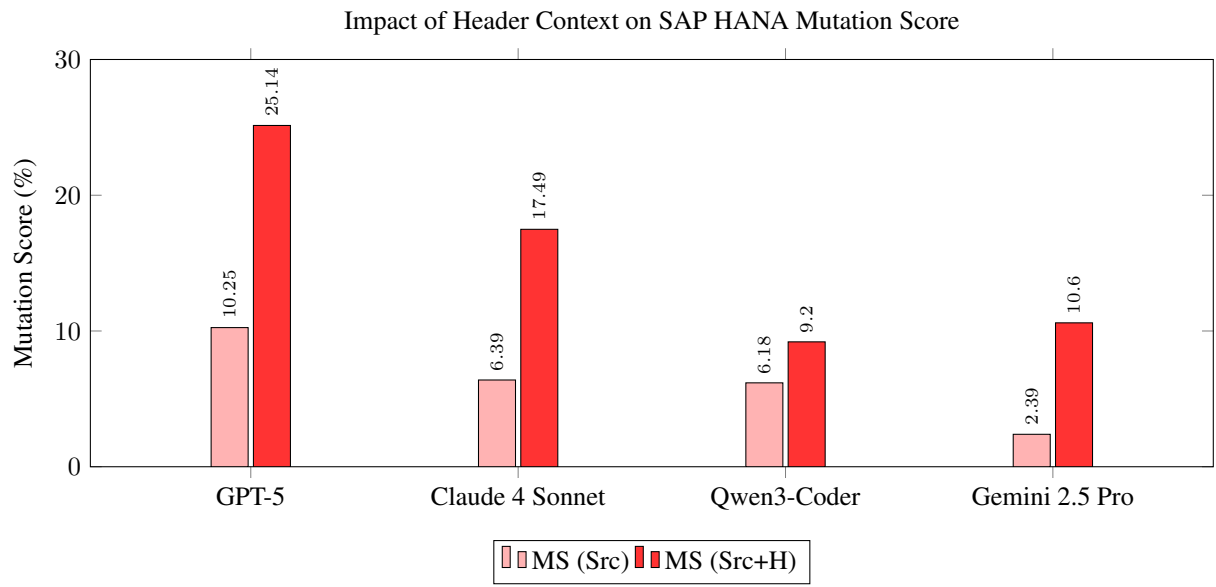
\begin{figure*}[t!]
\centering
{\begin{tikzpicture}
\begin{axis}[
    ybar,
    width=\textwidth,
    height=7cm,
    bar width=14pt,
    symbolic x coords={GPT-5, Claude 4 Sonnet, Qwen3-Coder, Gemini 2.5 Pro},
    xtick=data,
    ylabel={Mutation Score (\%)},
    ymin=0, ymax=30,
    legend style={at={(0.5,-0.18)}, anchor=north, legend columns=2},
    title={Impact of Header Context on SAP HANA Mutation Score},
    nodes near coords,
    nodes near coords style={font=\scriptsize, rotate=90, anchor=west},
    enlarge x limits=0.2
]

\addplot[fill=red!30] coordinates {(GPT-5,10.25) (Claude 4 Sonnet,6.39) (Qwen3-Coder,6.18) (Gemini 2.5 Pro,2.39)};
\addplot[fill=red!80] coordinates {(GPT-5,25.14) (Claude 4 Sonnet,17.49) (Qwen3-Coder,9.20) (Gemini 2.5 Pro,10.60)};

\legend{MS (Src), MS (Src+H)}
\end{axis}
\end{tikzpicture}}
\caption{Effect of header context on mutation score for SAP HANA~\citep{bekm26c}.}
\label{fig:hana_mutation_context}
\end{figure*}

Figures~\ref{fig:hana_coverage_context} and~\ref{fig:hana_mutation_context}
show the effect of header context on coverage and mutation score across
all four models on SAP HANA. The improvement is consistent across every
model, confirming that the knowledge gap is a systematic property of
the unseen codebase rather than a model-specific artifact~\citep{bekm26c}.

\section{Discussion}\label{discussion}

The combined findings from ARC-AGI evaluations and industrial unit test
generation demonstrate that current LLMs rely heavily on data
contamination and surface-level heuristics~\citep{mitc23a,mitc19a}. When
faced with novel, complex domains, whether generating test suites for
SAP HANA or solving abstract visual reasoning
puzzles~\citep{lewi24a}, models lack robust ``world models'' and instead
attempt to brute-force solutions.

The observation that LLMs generate trivial tests merely to satisfy the
compiler perfectly illustrates Goodhart's Law: when a measure
(e.g., compatibility or code coverage) becomes a target, it ceases to
be a good measure~\citep{good75a, stra97a}. The models successfully
reached the target of producing compiling code, but entirely missed the
underlying goal of generating tests capable of detecting software faults.
Similarly, in ARC-AGI, models may output the correct grid by relying on
shallow features like pixel counts, completely missing the intended
human-like abstraction.

What makes the iterative repair finding particularly notable is that the
shortcut-taking process is directly observable over time. With each
repair iteration on SAP HANA, models progressively resolved compilation
errors by removing assertions or generating empty test bodies rather than
by reasoning about the underlying code~\citep{bekm26c}. The metric
being optimized is compilability which diverged from the actual goal of fault
detection.

\paragraph{Implications.}

These results highlight the necessity of fundamentally changing how AI
is evaluated. Researchers must look beyond saturated benchmarks and
simple accuracy metrics, investigating the \emph{way} models reach their
answers to ensure they align with actual task competence.

In the software testing domain, this means supplementing
code coverage with mutation score as evaluation
metric~\citep{inoz14a,bekm26c}. Coverage measures
execution; mutation score measures comprehension. A model that achieves
high coverage through trivial tests is taking the same kind of shortcut
as a model that solves ARC-AGI through pixel counting where the output looks
correct, but the mechanism is not.

Future work should integrate rigorous semantic feedback mechanisms to
penalize trivial shortcuts and reward genuine fault-detection capability.
One concrete direction is to incorporate mutation analysis directly into
the repair loop: instead of feeding back only compiler errors, surviving
mutants could be fed back to the model as explicit generation
targets~\citep{bekm26c}. Preliminary evidence from related work suggests
that mutation-guided generation produces tests with substantially higher
fault-detection effectiveness~\citep{harm25a}.

Another direction concerns context and knowledge injection. The
performance improvements from providing dependency files suggest that
static context injection alone is insufficient for complex industrial
codebases. More promising are agentic approaches, where LLMs are
equipped with tools to autonomously retrieve relevant project context,
navigate dependency graphs, and iteratively interact with the build
environment~\citep{jime24a}. Such agentic pipelines could
dynamically assemble the minimal context required for each generation
task, without requiring project-specific fine-tuning, and are a natural
next step for test generation in large proprietary systems.

\paragraph{Limitations.}

Several limitations of the underlying study should be acknowledged.
The SAP HANA evaluation was restricted to a single representative
component due to the computational cost of mutation testing at industrial
scale~\citep{bekm26c}. Each experimental configuration was repeated only
twice, which is insufficient for formal statistical significance testing.
Additionally, LevelDB is considerably smaller and less architecturally
complex than SAP HANA: 21,207 lines of code versus 70,678 for the
evaluated component in SAP HANA, which means the two systems differ not only
in training data exposure but also in structural complexity, and this
distinction should be kept in mind when interpreting the performance
gap~\citep{bekm26c}. Finally, the mutation operator set was restricted
to arithmetic, boundary, comparison, and logical operators. A broader
operator set might reveal additional weaknesses in the generated tests.

However, we assume these limitations do not affect the core argument:
the performance contrast between LevelDB and SAP HANA is large enough to be robust to
moderate experimental variance, but they do constrain the
generalizability of the specific quantitative results.

\section{Conclusions}\label{conclusion}

This paper examined LLM shortcut behavior through the lens of automated
software test generation. By contrasting performance on a familiar
open-source system (LevelDB) with a proprietary industrial codebase
guaranteed to be absent from training data (SAP HANA), we documented
three concrete shortcut behaviors: performance collapse under
distribution shift consistent with memorization rather than
generalization, progressive degradation of test quality under iterative
repair as models optimize for compilability over fault detection, and a
near-complete absence of implicit structural knowledge on unseen
codebases.

Together, these findings provide empirical software engineering evidence
for the broader claim advanced by
Mitchell~\citep{mitc23a,mitc19a,bege25a} that current LLMs exploit
surface heuristics rather than demonstrating robust reasoning, and that
evaluation frameworks must examine mechanisms, not just outcomes. Mutation score, combined with an iterative repair loop,
offers a practical and principled way to make shortcut behavior
directly observable in code generation tasks.

These findings carry consequences that extend well beyond software
testing. Marcus~\citep{marc18a} has argued that deep learning suffers
from systematic limitations, including brittleness, superficial
generalization, and an inability to distinguish correlation from
causation, that no amount of scaling is likely to resolve. The shortcut
behaviors documented here provide independent, empirical confirmation of
precisely these limitations in a concrete engineering setting.
When LLMs are deployed in domains where their training data creates an
illusion of competence, whether in medical diagnosis, legal reasoning,
scientific analysis, or public decision-making, the same mechanisms
apply: surface-level pattern matching may produce plausible outputs that
lack the causal understanding required for reliable
results~\citep{bart19a}. The gap between memorized competence on LevelDB
and genuine reasoning failure on SAP HANA is not merely a software
engineering concern; it is a microcosm of the broader risk that
Mitchell~\citep{mitc19a}, Marcus~\citep{marc18a}, and
Bartz-Beielstein~\citep{bart19a} have identified from complementary
perspectives: society increasingly relies on systems whose fluent outputs
mask fundamental deficits in understanding.

We hope these results open a productive dialogue between cognitive
science and software engineering research communities on the shared
challenge of building evaluations that are resistant to the shortcuts
models have proven remarkably effective at finding.

\bibliographystyle{plainnat}
\bibliography{references}

\end{document}